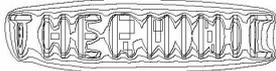

# IMPROVEMENTS OF THE VARIABLE THERMAL RESISTANCE

*V. Székely[1,2], S. Török[1], E. Kollár[1]*

[1]Budapest University of Technology & Economics
Department of Electron Devices
szekely|torok|kollar@eet.bme.hu
[2]MicReD Microelectronics Research and Development Ltd.

**ABSTRACT**

A flat mounting unit with electronically variable thermal resistance [1] has been presented in the last year [2]. The design was based on a Peltier cell and the appropriate control electronics and software. The device is devoted especially to the thermal characterization of packages, e.g. in dual cold plate arrangements.
Although this design meets the requirements of the static measurement we are intended to improve its parameters as the settling time and dynamic thermal impedance and the range of realized thermal resistance. The new design applies the heat flux sensor developed by our team as well [3], making easier the control of the device. This development allows even the realization of negative thermal resistances.

**Keywords:** variable thermal resistance, Peltier cell, heat-flux sensor, DCP measurements

## 1. INTRODUCTION

Let us briefly summarize the operating principle of the first version of the device. The electronically *variable thermal resistance* (VTR) mount is a sandwich structure, consisting of a Peltier cell, with a heat spreading plate in both sides. Temperature sensors are placed in both sides of the structure; the temperature data are forwarded to the control unit. This unit provides the driving current for the Peltier cell. The Peltier current is controlled by the two temperatures $T_1$ and $T_2$ in such a way that the $R_{thv}$ virtual thermal resistance "seen" on the top of the structure has to be the prescribed value.

The above realization has some drawbacks. Due to the control based on the $T_1$ and $T_2$ temperatures the settling times are high. The realized $R_{th}$ value is very sensitive to the exact values of Peltier cell parameters. Moreover, the stability of the control is problematic if we try to realize negative $R_{th}$ values.

In order to overcome these weaknesses we completed the mount by a heat-flux sensor. This paper is dealt with the experiences gained by this new structure.

## 2. THE NEW MOUNT STRUCTURE

The new mount structure is shown in Fig. 1. Two heat spreader metal layer and a Peltier cell constitute a sandwich structure (similarly to the former version). The temperatures of both metal layers are measured by diode temperature sensors. The improvement is that a heat flux sensor is placed on the top of the structure. Using this sensor the total heat flux penetrated at the top surface can be measured and used in the control algorithm. The heat flow sensor is a full-silicon, gradient type sensor, characterized by a very small thermal insertion resistance and good sensitivity. Design of this sensor is described in our earlier paper [3].

## 3. CONTROL HARDWARE AND ALGORITHM

The voltage of the temperature sensor diodes is amplified and converted to digital data. The measured range is 0…100 °C, the resolution (LSB) is 0.0015 °C. The output voltage of the heat flux sensor is very small (~10 µV/W) requiring an amplifier stage with voltage gain of about 20000. Due to the noise considerations the bandwidth of this amplifier is limited in only 10 Hz. The heat flux data is converted into digital one as well.

The control of the variable thermal resistance unit is realized by software. The $T_1$ and $T_2$ temperature data converted to digital are read in into the control program that runs on a PC. The program controls the current of the Peltier cell via an AD converter and a power voltage/current converter.

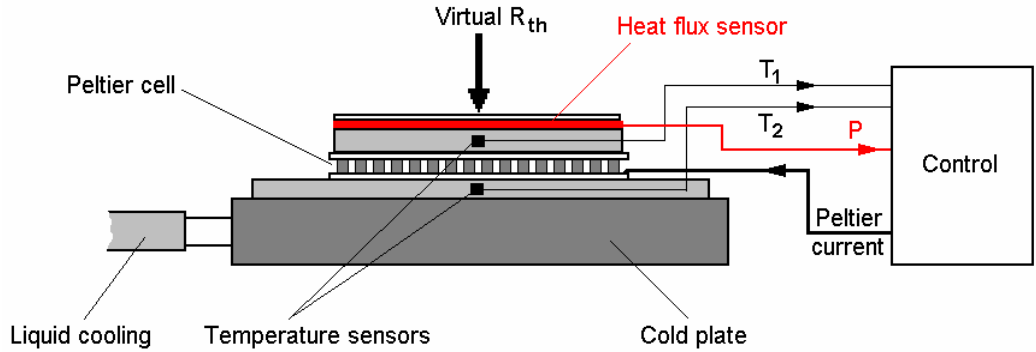

*Fig.1. The mount completed with a heat flux sensor*

The user has to declare two data for the control program:

- The required virtual thermal resistance $R_{thv}$
- The nominal ambient temperature $T_{amb}$ (practical but not obligatory if this latter is equal to the temperature of the cold plate).

The control software measures continuously the incoming heat flux $P$ and calculates the $T_1^*$ "required" temperature of the upper heat spreader sheet in the following way:

$$T_1^* = T_{amb} + P \cdot R_{thv} \qquad (1)$$

In order to calculate the current of the Peltier cell we apply proportional control, by using the equation:

$$I_{Peltier} = G(T_1^* - T_1) \qquad (2)$$

where the value of $G$ has to be determined considering the stability limit of the system. Visibly the $T_2$ temperature does not participate now the control algorithm but may serve the monitoring of the cold plate temperature.

4. EXPERIMENTAL RESULTS

In order to measure the parameters of the VTR unit, a bipolar transistor has been mounted on the top of the device, serving as a thermal impedance probe. This transistor was driven and measured by the T3Ster thermal transient tester equipment [4].

First the static behavior of the VTR unit has been examined. Some of the results are plotted in Fig. 2.

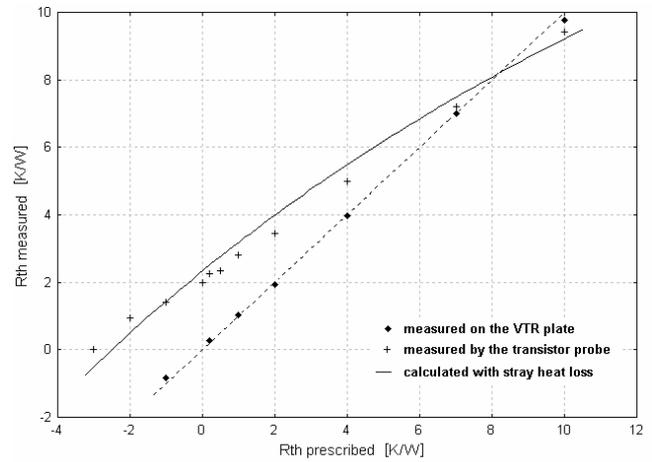

*Fig. 2. Measured vs. prescribed thermal resistance values*

This diagram shows the measured static $R_{th}$ values as the function of the declared (prescribed) $R_{th}$ data. The solid squares have been measured directly on the top of the device using $T_1$ while the crosses mark the data measured by the transistor probe. The latter involves the thermal resistance of the probe transistor and the upward direction heat loss. By taking into account these effects with 2.5 K/W and 35 K/W thermal resistance respectively, the solid line can be calculated.

The static behavior is well represented by the curves of Fig. 3 where the change of $T_1$ temperature is plotted as the function of the injected power. The different curves belong to different virtual thermal resistance values, between –2 K/W and 6 K/W. Linear behavior corresponds to straight lines in this diagram. Slight curvature of the diagrams indicates slight nonlinearity. The slope of the lines gives the realized thermal resistance values. These values are referred to the plain interface just below the heat flow sensor.

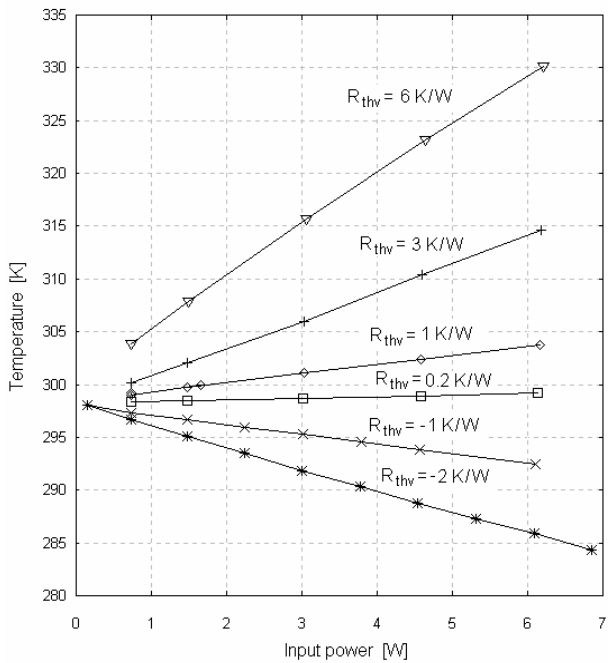

*Fig.3. Measured characteristics of the VTR unit, programmed to six different $R_{thv}$ values.*

The quasi-linear region of the VTR characteristics is limited by the finite cooling or heating capability of the

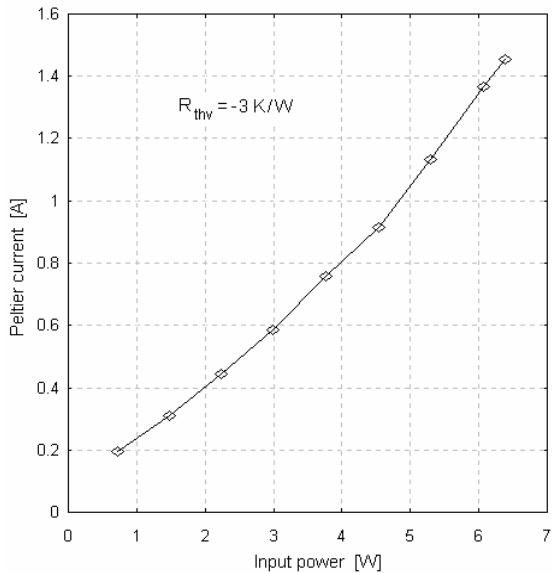

*Fig. 4. Peltier current vs. injected power for a negative (–3 K/W) $R_{thv}$ value*

Peltier cell. This limitation is stronger in case of negative $R_{th}$ values. Fig. 4 shows the steep increase of the current of Peltier cell, in case of $R_{thv}$ = -3 K/W. If the current of the Peltier cell is limited in e.g. 2 A, the –3 K/W can be realized up to 7.5 W injected power. For positive virtual thermal resistance values this limitation is much less severe.

In order to investigate the dynamic behavior of the unit thermal transients have been measured. The heating curves are plotted in Fig. 5 for five different values of the prescribed thermal resistance. It is obvious that the beginning parts of the curves are the same for all cases since this part corresponds to the thermal structure of the transistor probe. It should to draw the attention to the –1 K/W and –2 K/W curves where the temperature reaching a maximum value falls finally. This phenomenon can be explained as the effect of the realized negative thermal resistance.

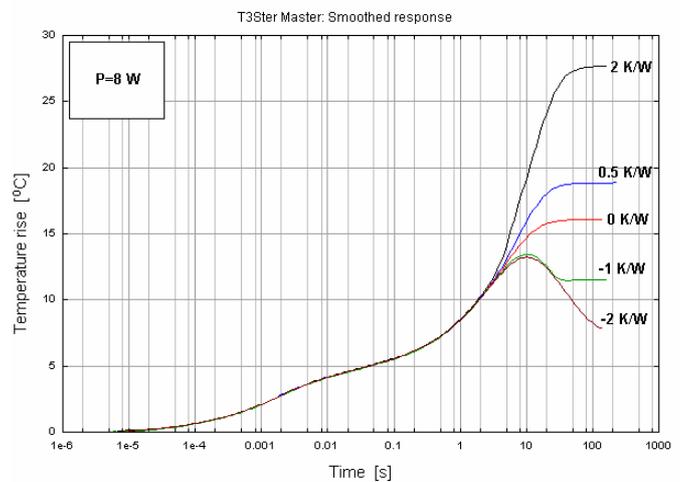

*Fig. 5. Heating curves for five programmed values of the virtual thermal resistance*

The heating curves can be transformed into Nyquist plots in the frequency domain. Such a plot is presented in Fig.6 for five prescribed $R_{th}$ values.

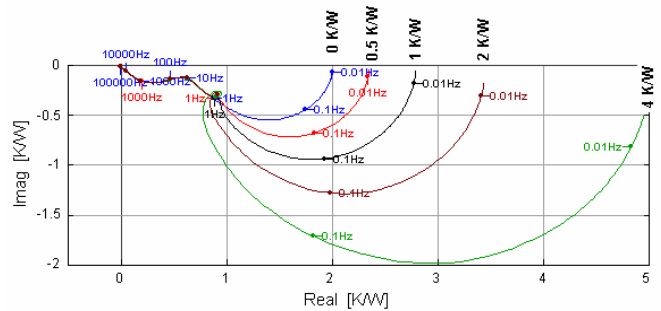

*Fig. 6. Loci of the complex thermal impedance (Nyquist plots) for five $R_{thv}$ values*

Again the beginning parts of the diagrams are coincident. This region corresponds to the internal thermal structure of the transistor probe. The remaining region of the curves is more or less a semi-circle suggesting that the VTR device can be characterized by a single RC stage. The τ time constants are about 5 s for 0 K/W and 19 s for 4 K/W.[1]

Finally the cumulative structure functions are presented (Fig.7). This diagram shows very clearly that the thermal resistance of one section of the heat flow path is variable, programmable by the control program of the VTR device.

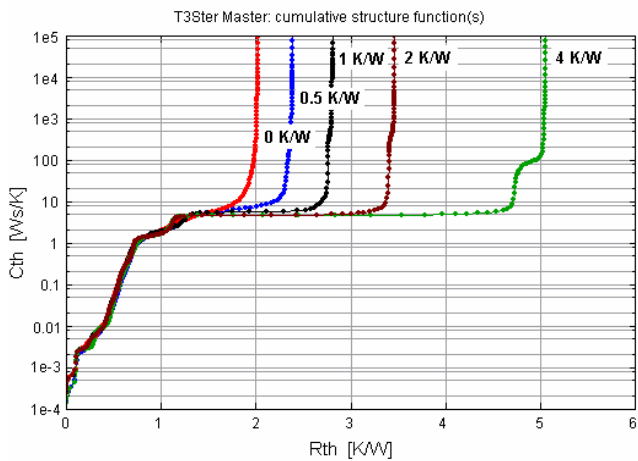

*Fig.7. Cumulative structure functions for five prescribed $R_{th}$ values*

## 5. CONCLUSIONS

By using a built-in heat-flux sensor the properties of the variable thermal resistance unit can be improved. The range of the realized $R_{th}$ values is broadened, even negative thermal resistance can be realized.

The time-dependent properties of the VTR unit can be described by a single thermal RC stage (by a single time constant).

## 6. ACKNOWLEDGMENTS

The authors wish to express their gratitude to M. Ádám for her contribution in the production of heat/flow sensors. The support of this work by the Hungarian grant "INNOCHECK" KM-CHECK-2005-00043 is acknowledged as well.

---

[1] The settling time for 99% of the final value is 4.6·τ.